 \documentclass[12pt, draftclsnofoot, onecolumn]{IEEEtran}
\usepackage{cite}
\usepackage{amsmath}
\usepackage{graphicx, subcaption}
\usepackage{subcaption}
\usepackage{float}
\floatstyle{plaintop}
\restylefloat{table}
\usepackage[justification=centering]{caption}
\usepackage{amssymb}
\usepackage{amsmath}
\usepackage{filecontents}
\usepackage{lipsum}
\usepackage{amsthm}
\usepackage{algorithmic,cite}
\usepackage{color}
\usepackage{authblk}
\usepackage[linesnumbered,ruled]{algorithm2e}
\usepackage{enumitem} 
\usepackage{booktabs}
\usepackage{chngcntr}

\counterwithin{subsubsection}{section}

\newtheorem{remark}{Remark}

\DeclareMathOperator{\Tr}{Tr}
\DeclareMathOperator{\diag}{\textbf{diag}}

\allowdisplaybreaks

\begin{document}
	\title{Machine Learning Methods for User Positioning With Uplink RSS in Distributed Massive MIMO}
	\author{K. N. R. Surya Vara Prasad, Ekram Hossain, and Vijay K. Bhargava. \thanks{K. N. R. S. V. Prasad and V. K. Bhargava are with the Department of Electrical and Computer Engineering at the University of British Columbia, Canada (emails: $\{$surya, vijayb$\}$@ece.ubc.ca). E. Hossain is with the Department of Electrical and Computer Engineering at the University of Manitoba, Canada (email: ekram.hossain@umanitoba.ca).}
		\thanks{Initial results of this work were presented at the 11th IEEE International Conference on Advanced Networks and Telecommunications Systems (IEEE ANTS), Bhubaneswar, India, December 2017 \cite{nagp}.}}
	\maketitle
	\begin{abstract}
		We consider a machine learning approach based on Gaussian process regression (GP) to position users in a distributed massive multiple-input multiple-output (MIMO) system with the uplink received signal strength (RSS) data. We focus on the scenario where noise-free RSS is available for training, but only noisy RSS is available for testing purposes. To estimate the test user locations and their $2\sigma$ error-bars, we adopt two state-of-the-art GP methods, namely, the conventional GP (CGP) and the numerical approximation GP (NaGP) methods. We find that the CGP method, which treats the noisy test RSS vectors as noise-free, provides unrealistically small $2\sigma$ error-bars on the estimated locations. To alleviate this concern, we derive the true predictive distribution for the test user locations and then employ the NaGP method to numerically approximate it as a Gaussian with the same first and second order moments. We also derive a Bayesian Cramer-Rao lower bound (BCRLB) on the achievable root-mean-squared-error (RMSE) performance of the two GP methods. Simulation studies reveal that: (i) the NaGP method indeed provides realistic $2\sigma$ error-bars on the estimated locations, (ii) operation in massive MIMO regime improves the RMSE performance, and (iii) the achieved RMSE performances are very close to the derived BCRLB.  
	\end{abstract}
\section{Introduction}
Wireless user positioning is an important research direction for the fifth generation (5G) networks because location information can be utilized to provide context-aware communication services. For example, approximate location information can facilitate area-specific advertisements, content caching, and also personnel tracking under emergency calls. Satellite-based global positioning systems (GPSs) \cite{gps_ref}, which are currently being used in LTE networks to procure location information, suffer from two major limitations. Firstly, GPSs provide unreliable location estimates for indoor and non-line-of-sight users. Secondly, GPS sensors are among the most power-hungry ones on a mobile device \cite{gps_ref2}, thus often causing the users to turn off their GPS functionality. These shortcomings have led to much research focus on alternative local positioning systems (LPSs), which use information available locally within the network, such as angle-of-arrival (AOA) and departure (AOD), time of arrival (TOA), and received signal strength (RSS) of the wireless signals, to position the users \cite{lps_survey1}. Out of these, RSS-based LPSs enjoy the advantage that no special measurement hardware needs to be installed at the BS.

The massive multiple-input multiple-output (MIMO) technology \cite{marzetta}, \cite{ee_paper}, which operates with a large number of antennas at the base station (BS), opens up new opportunities for the use of machine learning in LPS design. Due to the large number of BS antennas, massive MIMO allows the BS to record large vectors of signal properties, such as RSS, TOA, and AOA, whenever a user transmits signals on the uplink. Machine learning techniques can then be employed for user positioning, wherein we choose a signal property (for example, TOA) and train a machine learning model with a database comprising of signal property vectors recorded at several known user locations. The trained machine learning model is then used to predict the location of a test user when its signal property vector is provided as the test input. 

In this work, we consider a distributed massive MIMO \cite{dm_mimo} setup and propose a new machine learning approach based on Gaussian process regression (GP) to predict user locations from their uplink RSS data. We rely on the GP framework to build our machine learning approach because GP allows us to derive location estimates in the form of a full predictive distribution \cite{gpr_book}, i.e., we can obtain closed-form expressions for the predicted locations and the associated $2\sigma$ error-bars. 
\color{black}We choose RSS as the signal property for user positioning because RSS measurements are readily available at the BS, without the need for extra hardware to be installed. One of the major difficulties with using RSS for user positioning is that the RSS data is generally corrupted with noise due to small-scale fading and shadowing effects of the wireless channel. Small scale fading can be mitigated by averaging over multiple timeslots and subcarriers, but it is difficult to mitigate shadowing because spatial averaging, which requires prior knowledge of the user location, should be employed \cite{posn1} \cite{posn2}. Since we are not aware of the test users' locations, we cannot average out the shadowing noise present in the test RSS data. In contrast, we can synthetically generate noise-free RSS data for training purposes. To do so, we only require knowledge of the BS antenna locations, the training user locations, the uplink transmission power, and the path-loss exponent for the area of operation.

With the above constraint in mind, we investigate user positioning in the scenario where the training RSS data is noise-free and the test RSS data is noisy due to shadowing effects. Firstly, we consider the conventional GP (CGP) method, which naively treats the test RSS data as noise-free for location prediction. Our simulation studies reveal that the CGP method provides unrealistically small $2\sigma$ error-bars on the predicted locations. To address this limitation, we consider the use of a moment-matching based GP method, referred to in this work as the numerical approximation GP (NaGP) method, for location prediction. The NaGP method derives the true predictive distribution of the test user locations by learning from the statistical properties of the noise present in the test RSS. Since the true predictive distribution cannot be obtained in a closed-form, it is approximated numerically as a Gaussian distribution with the same first and second order moments. While the first order moments give us the location estimates for the test users, the second-order moments give us realistic estimates of the associated $2\sigma$ error-bars. The main contributions of our work are summarized as follows
\begin{itemize}
	\item[(i)] For the machine learning problem of RSS-based user positioning in distributed massive MIMO, ours is the first work to identify that the conventional GP (CGP) method provides unrealistically small $2\sigma$ error-bars on the predicted locations. To derive realistic $2\sigma$ error-bars on the location estimates, we are also the first to apply the NaGP method, which is a popular moment-matching-based GP method in time-series analysis \cite{gplit1}. The NaGP method derives the true predictive distribution and then approximates it as a Gaussian distribution with the same first and second order moments, so as to yield the location estimates and their $2\sigma$ error-bars. 
	\item[(ii)] Unlike most existing machine learning approaches, we derive closed-form expressions for the estimated locations and their $2\sigma$ error-bars in terms of the training RSS, the training user locations, and the test RSS data. 
	\item[(iii)] We derive a Bayesian Cramer-Rao lower bound (BCRLB) on the achievable RMSE performance of the two GP methods under study. Our derivation reveals that the BCRLB can be computed via simple linear algebraic operations on the obtained predictive variances.
	\item[(iv)] We demonstrate the benefit of massive MIMO from the perspective of RSS-based user positioning. We report that when the training RSS is noise-free and the test RSS is noisy due to shadowing, an increase in the number of base station antennas beyond the conventional MIMO standards can result in an improved root-mean-squared error performance.

\end{itemize}

The rest of the paper is organized as follows. In Section \ref{sec_related_work}, we present a review of existing works on the topic of study. In Section \ref{sec_sys_model}, we present a distributed massive MIMO setup and explain how machine learning can be employed for user positioning. In Section \ref{sec_train}, details are presented on the training phase that is common to the two GP methods under study. The CGP and NaGP methods are presented in Sections \ref{sec_sgp}-\ref{sec_nagp_gagp}. Performance metrics are presented in Section \ref{sec_perf} and a Bayesian Cramer-Rao lower bound is derived on the achievable RMSE performance. Numerical results are presented in Section \ref{sec_sim} to validate the prediction performance of the two GP methods, followed by few concluding remarks in Section \ref{sec_conclusion}.

\color{black}
\section{Related Work} \label{sec_related_work}
\subsection{User Positioning in Massive MIMO} Most research works on user positioning in massive MIMO have been very recent. The authors in \cite{mmloc1} propose a compressed sensing approach to estimate user locations from TOA information recorded at multiple massive MIMO BSs. An optimization problem is solved over a convex search space formed by coarse TOA estimates recorded at each BS, so as to estimate the user location. AOA information is used in \cite{mmloc2}-\cite{mmloc4}, while the combined information of time delay, AOD, and AOA information is used in \cite{mmloc5} for positioning users in massive MIMO. A millimeter wave massive MIMO system is considered in \cite{mmloc4} to derive the necessary conditions under which a user location can be estimated from AOD and AOA information under line-of-sight conditions. 

The most related work to our study is \cite{rss_gp5}, where RSS-based user positioning is investigated, but noisy RSS data is considered for both training and prediction. The GP method proposed in \cite{rss_gp5} does not learn from the noisy nature of the test RSS data. In contrast,  we study the scenario where the training RSS data is noise-free and only the test RSS data is noisy due to shadowing. Similar to \cite{rss_gp5}, we consider using the conventional GP method \cite{gpr_book} for location prediction. But in addition, we identify that the CGP method provides unrealistically small $2\sigma$ error-bars on the estimated location. To address this concern, we apply a moment-matching based GP method, namely the NaGP, to position users from their RSS data in a distributed massive MIMO setup. The NaGP method derives and approximates the true predictive distribution, so as to provide more realistic $2\sigma$ error-bars on the location estimates than the CGP method. 
\subsection{Machine Learning Techniques for User Positioning}
Several machine learning  techniques, including, neural networks \cite{mllit2}, $k$-nearest neighbours \cite{mllit3}, support vector machines 
\cite{mllit4}, \cite{mllit5}, GP methods \cite{rss_gp5}, and more recently, deep learning methods \cite{DL1}-\cite{DL3}, have been explored for user positioning in a variety of wireless networks. 
\color{black} From these techniques, we choose the GP framework for our analysis for three reasons. Firstly, GP methods are known to perform as good as most other machine learning methods \cite{gpw1}\cite{ras_thesis}. Secondly, unlike other machine learning methods, GP methods provide probabilistic location estimates in the form a full predictive distribution. This allows us to obtain closed-form expressions for both the predicted locations and the associated 2$\sigma$ error-bars. Thirdly, most of the explored machine learning methods, including the recent deep learning techniques \cite{DL1}-\cite{DL3}, do not lend themselves to rigorous performance guarantees \cite{DL4}. In contrast, GP methods allow us to derive Cramer-Rao type bound on the prediction performance.

\color{black}
Coming to GP methods for user positioning in wireless networks, we observe that most of the existing works \cite{gpw1}-\cite{rss_gp4} have opted for an indirect modelling approach, wherein the GP models take location estimates as inputs and provide RSS values as outputs. User locations are then obtained by maximizing the joint likelihood of the output RSS values. Following this approach, a GP method is proposed in \cite{gpw1} to position indoor users based on downlink RSS from multiple BSs. Each user trains one GP model per BS and predicts its own location by maximizing the joint likelihood of the downlink RSSs. Since the joint likelihood can be highly peaked, a smoothing procedure is proposed in \cite{rss_gp2}. Both \cite{gpw1} and \cite{rss_gp2} impose the non-trivial task of choosing appropriate initial values for the likelihood maximization problem. In this regard, \cite{rss_gp3} proposes a GP method which trains an extra GP model per BS, so as to obtain raw location estimates for initialization. Similar to \cite{gpw1}-\cite{rss_gp2}, a GP method is proposed in \cite{rss_gp4}, but for localization in WiFi networks. Lower bounds are derived for the estimation error on the free parameters introduced by the GP model. Our work is different from the above works in three ways.

Firstly, all of the above works advocate the use of conventional GP methods for location prediction, wherein, the proposed methods do not formally utilize the statistical knowledge of the noise present in the inputs to improve the prediction performance. In contrast, we advocate the use of a moment-matching based GP method, namely the NaGP, for user positioning. The NaGP method performs better than the conventional GP method by learning from the statistical properties of the noise present in the inputs. Secondly, all of the above works take an indirect modelling approach, wherein, each user trains at least one GP per BS. Such a training policy may not be computationally feasible for users in a distributed massive MIMO setup because the system operates with a large number of remote radio heads. We therefore opt for a direct modelling approach, wherein, the GP model takes RSS values as inputs and provides location estimates as outputs. Our approach only requires two GP models to be trained in total - one for predicting the $x-$coordinates of the users and the other for $y$ coordinates. Lastly, in all of the above works, the users are burdened with all the computational load required for training and prediction. In our work, the BS handles this computational load, which is more appealing because user devices operate with limited battery power. 
\subsection{GP Methods With Noisy Inputs}
The problem of dealing with noisy inputs in GP has been investigated before. Recent works, including \cite{gplit4}-\cite{gplit10} and references therein, deal with noisy input GPs, but consider noisy inputs for both training and prediction. The authors in \cite{gplit1},\cite{gplit2} propose moment-matching based GP methods, which learn from statistical properties of the noise present in the test inputs to derive (and approximate) the true predictive distribution, for multi-step prediction in time series analysis. These methods have been adapted for system identification in \cite{gplit4} and for spatial wireless channel prediction in \cite{gplit7} \cite{gplit10}, but not for RSS-based user positioning in distributed massive MIMO. Also, unlike the above works, we derive a Cramer-Rao lower bound on the achievable RMSE performance of the employed GP methods. 
\color{black}

\subsubsection*{General Notation} We use regular font small letters for scalars, boldface small letters for vectors, and boldface capital letters for matrices, for example, $a$, $\mathbf{a}$, and $\mathbf{A}$, respectively. The notations $[\mathbf{a}]_i$, $[\mathbf{A}]_i$, and $[\mathbf{A}]_{ij}$ refer to the element $i$ in vector $\mathbf{a}$, column $i$ in matrix $\mathbf{A}$, and the element $(i,j)$ in matrix $\mathbf{A}$, respectively. The overhead symbol $\widetilde{(.)}$ refers to training data and the overhead symbol $\widehat{(.)}$ refers to test data, respectively. An additional superscript $(.)^*$  is used if the data is noise-free.
The symbol $\approx$ denotes that we approximate the left hand side with the right hand side. The notation $\Tr(\mathbf{A})$ refers to the trace of the matrix $\mathbf{A}$. A random vector $\mathbf{a}$ that is Gaussian distributed with mean $\mathbf{u}$ and covariance $\mathbf{A}$ is referred to as $\mathbf{a}\sim\mathcal{N} (\mathbf{u}, \mathbf{A})$, and its probability density function (pdf) is denoted as $\mathcal{N} ({\mathbf{a}}; \mathbf{u}, \mathbf{A})$. Lastly, when ${\mathbf{u}}$ and $\mathbf{a}$ are deterministic $n$-dimensional vectors and $\mathbf{A}$ is a deterministic $n \times n$ matrix, we use the notation $N ({\mathbf{a}}; \mathbf{u}, \mathbf{A})$ as a shorthand for the expression $\{(2 \pi)^{-n/2} $ $ |\mathbf{A}|^{-1/2} e^{ -\frac{1}{2}(\mathbf{a} - \mathbf{u})^T \mathbf{A}^{-1} (\mathbf{a} - \mathbf{u})} \}$.

\color{black}
\section{System Description} \label{sec_sys_model}
\begin{figure}[h] 
\begin{center}
 \vspace{-3mm}
 \includegraphics[scale=0.45] {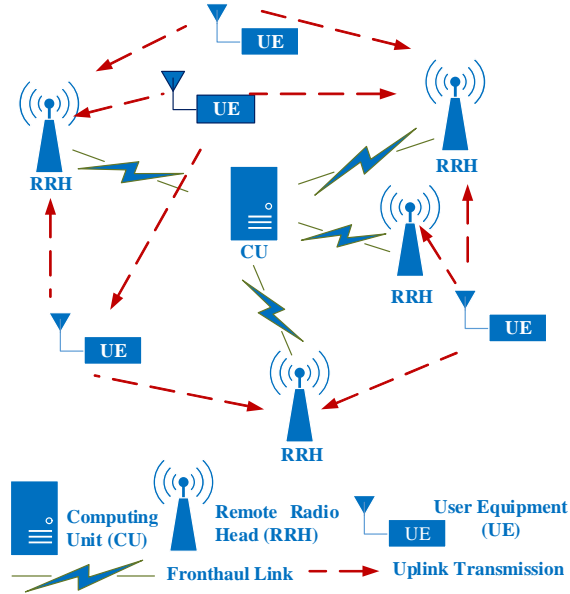} 
 \caption {Setup for user positioning in distributed massive MIMO: $K$ single-antenna UEs transmit uplink signals simultaneously to $M$ RRHs on the same time-frequency resource. Each RRH records its own RSS value and forwards it to the CU via high-speed fronthaul. The CU hosts a machine learning model which takes the RSS vectors as input so as to predict the transmitting user's location.  
 \label{fig_dmimo}}
 \end{center}
  \vspace{-3mm}
 \end{figure}

We consider a distributed multiuser massive MIMO setup, as shown in Fig. \ref{fig_dmimo}, where $K$ user equipments (UEs)  transmit uplink radio signals to $M$ remote radio heads (RRHs) simultaneously on the same time-frequency resource. For simplicity, we assume that the RRHs are all single-antenna units and refer to them interchangeably as BS antennas. We also assume that all the UEs in the system are single-antenna units and refer to them as users. The RRHs are connected to a central computing unit (CU) through high-speed fronthaul links. When the $K$ users transmit radio signals on the uplink simultaneously, each RRH records its own received signal strength (RSS). The CU gathers the recorded RSS values from each RRH, processes them to extract the per-user RSS values, and forms an $M \times 1$ RSS vector for each user. The RSS vectors thus formed are fed as input to a trained machine learning model for predicting the locations of the transmitting users. The CU hosts the machine learning model and handles all the computations required for the training and prediction. Details on the multiuser transmissions, per-user RSS extraction, and the mathematical model for machine learning are presented next.

\subsection{Multi-user Transmissions} 
When the user $k$ transmits a symbol vector $\boldsymbol{\omega}_k$ with power $\rho$, the BS antenna $m$ receives a symbol vector $\mathbf{r}_m$, given by \begin{equation} \label{eqRxsignal}
\mathbf{r}_m = \sqrt{\rho} \sum_{k=1}^{K} h_{mk} \boldsymbol{\omega}_k + \boldsymbol{\vartheta}_m,
\end{equation}
where $h_{mk} = q_{mk} \sqrt{g_{mk}} $ is the flat-fading uplink channel gain with $q_{mk}$ and $g_{mk}$ being the small-scale and large-scale fading coefficients, and $\boldsymbol{\vartheta}_m \sim \mathcal{N} (\mathbf{0}, \sigma_{\vartheta}^2 \mathbf{I})$ is the additive white Gaussian noise vector. We assume that the small-scale fading coefficients $q_{mk}$ are independent and identically distributed (i.i.d) complex Gaussian random variables, i.e., $q_{mk} \sim \mathcal{CN} (0,1)$, and model the large-scale fading coefficient $g_{mk}$ as
\begin{equation} \label{eqChannel}
\begin{aligned}
g_{mk} = b_0 d_{mk}^{-\eta} 10^{\frac{z_{mk}}{10}},
\end{aligned}
\end{equation}
where $d_{mk}$ is the distance between the user $k$ and BS antenna $m$, $b_0$ is the path-loss at a reference distance $d_0$, $\eta$ is the path-loss exponent, and $z_{mk} \sim \mathcal{N} (0,\sigma_{z}^2 )$ is the channel gain due to shadowing noise. 

\subsection{Extracting Per-user RSS Values} 
From (\ref{eqRxsignal}), we note that the RSS $||\mathbf{r}_m||^2$ at RRH $m$ corresponds to the multiuser RSS because the received vector $\mathbf{r}_m$ is the sum of symbol vectors received from all the $K$ users. We cannot directly use the multiuser RSS $||\mathbf{r}_m||^2$ to position any given user $k$ because we would then be unable to distinguish among the $K$ users that are transmitting simultaneously. Instead, the RRH $m$ should extract the per-user RSS $p_{mk}$ of each user $k$ from $\mathbf{r}_m$ and use it for positioning the user $k$. This can be done if the symbol vectors $\{\boldsymbol{\omega}_k\}$ in (\ref{eqRxsignal}) are mutually orthogonal and are already known at the RRH, for example, $\{\boldsymbol{\omega}_k\}$ can be pilot sequences transmitted for channel estimation \cite{cf_mimo}. The RSS $p_{mk}$ of user $k$ can then be obtained from $\mathbf{r}_m$ as  
\begin{equation} \label{eq_pmk}
p_{mk} = \rho g_{mk} |q_{mk}|^2.
\end{equation}
Observe from (\ref{eq_pmk}) that the extracted per-user RSS values can be noisy due to small-scale fading and shadowing effects of the wireless channel. We assume that the small-scale fading is averaged out over multiple time-slots and focus on the scenario where only the shadowing noise exists. We do so because shadowing is space-dependent and requires access to the user location in order to be averaged out. The resulting RSS, after substituting the large-scale fading model in (\ref{eqChannel}), is given in dB scale as
\begin{equation} \label{eq_rss_dB}
\begin{aligned}
p_{mk}^{\text{dB}} & = p_0^{\text{dB}} - 10 \eta \log_{10} (d_{mk}) + z_{mk},
\end{aligned}
\end{equation}
\noindent where $p_0^{\text{dB}} = 10 \log_{10} (\rho b_0) $ is the uplink RSS at the reference distance $d_0$.
For each user $k$, the CU can then form an $M \times 1$ RSS vector $\mathbf{p}_k$ such that $[\mathbf{p}_k]_m = p_{mk}^{\text{dB}}$, i.e.,
\begin{equation} \label{eq_rss_vector}
\mathbf{p}_k = [p_{1k}^{\text{dB}} \,\,\, p_{2k}^{\text{dB}} \,\,\, \dots \,\,\, p_{Mk}^{\text{dB}}]^T.
\end{equation}

\subsection{Machine Learning Model} 

Let us define $f_x(.)$ and $f_y(.)$ as the functions which map the RSS vector $\mathbf{p}_k$ of any user $k$ in the system to its 2D location coordinates $(x_k,y_k)$, such that
\begin{equation} \label{eq_mapping}
\begin{aligned}
x_k = f_x(\mathbf{p}_k) \, \text{ and } \, y_k = f_y(\mathbf{p}_k) \quad \forall x_k, y_k.
\end{aligned}
\end{equation}
We employ supervised machine learning to learn the functions $f_x(.)$ and $f_y(.)$, wherein we first train a machine learning model with RSS vectors for several known user locations. The trained model is then fed with RSS vectors of test users as inputs, so as to obtain their location estimates. We consider noise-free RSS vectors for training because they are easy to generate. For this, we only need knowledge of the RRH locations, the training user locations, the uplink transmission power $\rho$, and the path-loss exponent $\eta$. In case the $\eta$ value is not available, we can conduct field measurements to record the training RSS and our {\em a priori} access to the training user locations can aid us in spatially averaging out the shadowing effects. As an example, for a given training user location, we can record multiple RSS readings at nearby locations with approximately the same BS-to-user distance and note the average of these readings as the training RSS vector for that location. Such an averaging mitigates the shadowing noise present in the training RSS vectors, making them noise-free. On the other hand, we treat the test RSS vectors as noisy due to shadowing because we are unaware of the test user's locations and are therefore unable to spatially average out the shadowing effects.

To predict the test user locations from their uplink RSS vectors, we adopt two Gaussian process regression methods from time-series analysis \cite{gpr_book} \cite{gplit1}, namely, the conventional GP (CGP) and the numerical approximation GP (NaGP) methods.
\color{black} Both the GP methods employ the same training procedure, but differ in terms of how the test user locations are predicted. Therefore, details on the training phase are presented first\footnote{For simplicity, we focus on the training and prediction of $x-$coordinates only, but the proposed machine learning procedure is applicable for prediction of $y-$coordinates as well.}. 
 
\section{Training Phase of the GP Methods} \label{sec_train}
At the core of all the Gaussian process regression methods is the assumption that the function to be learned, i.e., $f_x(.)$, is drawn from a zero-mean Gaussian process prior specified by a user-defined covariance function $\phi(.\, ,.)$  \cite{gpr_book}. This means that any finite number of $f_x(.)$ realizations are assumed to follow a joint Gaussian distribution with mean zero and covariance $\mathbf{\Phi}$, whose elements are given by the function $\phi(.\,,.)$. We refer to this assumption as
 \begin{equation} \label{eq_gp_prior}
f_x(.) \sim \mathcal{GP} (0, \phi(.\,,.)).
 \end{equation} 
 
\noindent The function $\phi(.\,,.)$ models the covariance of $x-$coordinates of any two users in the system as a function of their RSS vectors. We choose $\phi(.\,,.)$ as the weighted-sum of squared-exponential (SE), inner product (IP) and delta functions, given for any two RSS vectors $\mathbf{p}_k$ and $ \mathbf{p}_{k'}$, by
\begin{equation}\label{eq_cov_model}
\begin{aligned}
\phi(\mathbf{p}_k, \mathbf{p}_{k'}) =& \, \alpha e^{-\frac{1}{2} (\mathbf{p}_k - \mathbf{p}_{k'})^T \mathbf{B}^{-1} (\mathbf{p}_k - \mathbf{p}_{k'}) } + \gamma \mathbf{p}_k^T \mathbf{p}_{k'} + \sigma_{er}^2 \delta_{kk'}, \\
& \text{{where  $\mathbf{B} = \diag\{\beta_m\}, m = 1, \dots, M$,} and}  \\
& \delta_{kk'} = \{ 1 \text{ if } k = k', \, 0 \text{ if otherwise} \}.
\end{aligned}
\end{equation}
In (\ref{eq_cov_model}), the SE term $\alpha e^{-\frac{1}{2} (\mathbf{p}_k - \mathbf{p}_{k'})^T \mathbf{B}^{-1} (\mathbf{p}_k - \mathbf{p}_{k'}) }$ captures the dependence of $\phi(\mathbf{p}_k, \mathbf{p}_{k'})$ on the distance between the RSS vectors $\mathbf{p}_k$ and $\mathbf{p}_{k'}$. The IP term $\gamma \mathbf{p}_k^T \mathbf{p}_{k'}$ captures the dependence of $\phi(\mathbf{p}_k, \mathbf{p}_{k'})$ on the actual RSS vectors $\mathbf{p}_k$ and $\mathbf{p}_{k'}$. The delta term $\sigma_{er}^2 \delta_{kk'}$ captures the variance due to measurement errors in the $x-$coordinates and is typically known apriori. The parameters $\alpha$ and $\gamma$ in (\ref{eq_cov_model}) govern the overall variance of the $x-$coordinate function $f_x(.)$. Diagonal elements $\beta_m$ of the matrix $\mathbf{B}$ govern the distance to be moved along each dimension $m = 1, \dots, M$ of the RSS space until the function realizations $f_x(\mathbf{p}_k)$ and $f_x(\mathbf{p}_{k'})$ become uncorrelated.

Our objective in the training phase is to train a GP model to learn the $x-$coordinate function $f_x(.)$. To learn $f_x(.)$, it is sufficient to learn the free parameters introduced by the covariance model in (\ref{eq_cov_model}) because, from the GP assumption (\ref{eq_gp_prior}), we know that $f_x(.)$ is fully specified by the covariance function $\phi(.\,,.)$. Let us accumulate the free parameters in (\ref{eq_cov_model}) into an $(M + 2) \times 1$ vector $\boldsymbol{\theta}$ as
\begin{equation} \label{eq_theta}
\text{{$\boldsymbol{\theta} = [\alpha \,\, \beta_1 \,\, \dots \,\, \beta_M \,\, \gamma]^T$}}. 
\end{equation}
Let us assume that we have access to $\widetilde{L}$ training locations. We now introduce the $\widetilde{L} \times 1$ vector $\widetilde{\mathbf{x}}$ of training $x-$coordinates and the corresponding $\widetilde{L} \times M$ matrix $\widetilde{\mathbf{P}}$ of noise-free RSS training vectors, defined as follows
\begin{equation} \label{eq_ptrain}
\begin{aligned}
\widetilde{\mathbf{x}} = [\widetilde{x}_1 \,\,\, \widetilde{x}_2 \,\,\, \dots \, \widetilde{x}_{\widetilde{L}}]^T, \\
\widetilde{\mathbf{P}} = [\widetilde{\mathbf{p}}_1 \,\,\, \widetilde{\mathbf{p}}_2 \,\,\, \dots \widetilde{\mathbf{p}}_{\widetilde{L}}]^T,
\end{aligned}
\end{equation}
where the row $l$ in $\widetilde{\mathbf{P}}$, i.e., the training RSS vector $\widetilde{\mathbf{p}}_l$,  corresponds to the training $x-$coordinate $\widetilde{x}_l$, $\forall$ $l = 1, \dots, \widetilde{L}$.
\color{black} 
Since we have from (\ref{eq_mapping}) that the training $x-$coordinates consitute a finite set of $f_x(.)$ realizations over the training RSS vectors in $\widetilde{\mathbf{P}}$, we know from (\ref{eq_gp_prior}) that the training $x-$coordinates are jointly Gaussian distributed as
\begin{equation} \label{eq_trainingdist}
\begin{aligned}
\widetilde{\mathbf{x}} | \widetilde{\mathbf{P}}, \boldsymbol {\theta} & \sim \mathcal{N} ( \mathbf{0}, \widetilde{\mathbf{\Phi}}), \text{ where} \\
[\widetilde{\mathbf{\Phi}}]_{ll'} &= \phi(\widetilde{\mathbf{p}}_l, \widetilde{\mathbf{p}}_{l'}), \forall \, l, l' = 1, \dots, \widetilde{L}. 
\end{aligned}
\end{equation}
Eq. (\ref{eq_trainingdist}) gives us the log-likelihood expression of $\widetilde{\mathbf{x}}| \widetilde{\mathbf{P}}, \boldsymbol {\theta}$. We can now learn the vector $\boldsymbol{\theta}$ via maximum-likelihood as

\begin{equation}\label{eq_theta_opt}
\begin{aligned}
	\bar{\boldsymbol{\theta}} & = \underset{\boldsymbol{\theta}}{\arg \max} \log (p(\widetilde{\mathbf{x}} | \widetilde{\mathbf{P}}, \boldsymbol{\theta})), 
\end{aligned}
\end{equation}
where $\bar{\boldsymbol{\theta}} $ is the learned parameter vector. The optimization problem in (\ref{eq_theta_opt}) is non-convex, but can be solved for a local optimum using gradient ascent methods,  such as conjugate gradient and L-BFGS \cite{optim_book}, because we can obtain the first-order gradients with respect to $\boldsymbol{\theta}$ in closed-form. In this work, we use the conjugate gradient method \cite{optim_book} to obtain a local optimum vector $\bar{\boldsymbol{\theta}} $. Solving (\ref{eq_theta_opt}) for $\bar{\boldsymbol{\theta}}$ completes the training phase because the covariance function $\phi(.\,,.)$ fully specifies the unknown mapping function $f_x(.)$.

 In the prediction phase, let there be $\widehat{L}$ test users whose location coordinates need to be predicted. We now introduce the $\widehat{L} \times 1$ vector $\widehat{\mathbf{x}}$ of the test user $x-$coordinates, which needs to be predicted from an $\widehat{L} \times M$ matrix $\widehat{\mathbf{P}}$ of the noisy test RSS vectors, defined such that
	\begin{equation} \label{eq_ptest}
	\begin{aligned}
	\widehat{\mathbf{P}} = [\widehat{\mathbf{p}}_1\,\,\, \widehat{\mathbf{p}}_2 \,\,\, \dots \widehat{\mathbf{p}}_{\widehat{L}}]^T, \\
	\widehat{\mathbf{x}} = [\widehat{x}_1 \,\,\, \widehat{x}_2 \,\,\, \dots \,\,\, \widehat{x}_{\widehat{L}}]^T, 
	\end{aligned}
	\end{equation}
	where the RSS vector $\widehat{\mathbf{p}}_l$ corresponds to the test user whose $x-$coordinate is $[\widehat{\mathbf{x}}]_l = \widehat{x}_l$, $\forall = 1, \dots, \widehat{L}$.
	\color{black}
In the next two sections, we present details on the CGP and NaGP methods, with focus on their prediction phase only because the training procedure is the same as detailed above.
\section{Location Prediction With Conventional GP Method (CGP)} \label{sec_sgp}
We now consider CGP - a GP method which employs conventional GP principles\cite{gpr_book} to predict the user locations. The CGP method naively treats the noisy test RSS vectors as noise-free and uses the assumption (\ref{eq_gp_prior}) to obtain the joint distribution of the training and test $x-$coordinate vectors $\widetilde{\mathbf{x}}$ and $\widehat{\mathbf{x}}$ as
\begin{align} \label{eq_joint_dist}
\begin{bmatrix}
\widetilde{\mathbf{x}}\\
\widehat{\mathbf{x}}
\end{bmatrix} \Bigg{\vert} \widetilde{\mathbf{P}}, \widehat{\mathbf{P}}
  \sim  \mathcal{N}
\begin{bmatrix}
\begin{pmatrix}
\mathbf{0}\\
\mathbf{0}
\end{pmatrix}\!\!,
\begin{pmatrix}
\widetilde{\mathbf{\Phi}} & (\mathbf{\Phi}^{\dagger})^T\\
\mathbf{\Phi}^{\dagger} & \widehat{\mathbf{\Phi}} 
\end{pmatrix}
\end{bmatrix}, 
\end{align}
where $\widetilde{\mathbf{\Phi}} \in \mathbb{R}^{\widetilde{L} \times \widetilde{L}}$, $\mathbf{\Phi}^{\dagger} \in \mathbb{R}^{\widetilde{L} \times \widehat{L}}$, and $\widehat{\mathbf{\Phi}} \in \mathbb{R}^{\widehat{L} \times \widehat{L}}$ are the covariance matrices between the noise-free training and noisy test RSS vectors, defined such that 
\begin{equation} \label{eq_cov_matrices}
\begin{aligned}
&\text{$[\widetilde{\mathbf{\Phi}}]_{ll''}$} = \phi(\widetilde{\mathbf{p}}_l, \widetilde{\mathbf{p}}_{l'}), \quad l, l' = 1, \dots, \widetilde{L} \\
 & [\mathbf{\Phi}^{\dagger}]_{ll'}  = \phi(\widehat{\mathbf{p}}_{l}, \widetilde{\mathbf{p}}_{l'}), \quad l = 1, \dots, \widehat{L} , \quad l' = 1, \dots, \widetilde{L} \text{ and} \\
& [\widehat{\mathbf{\Phi}}]_{ll'} \, \, = \phi(\widehat{\mathbf{p}}_{l}, \widehat{\mathbf{p}}_{l'}), \quad l, l' = 1, \dots, \widehat{L}.
\end{aligned}
\end{equation}
Conditioning the joint distribution in (\ref{eq_joint_dist}) over $\widetilde{\mathbf{x}}$ gives us the posterior predictive distribution of $\widehat{\mathbf{x}}$ as (c.f. (\ref{eq_conditioning_formula}) in \textbf{Appendix})
\begin{equation} \label{eq_pred_sgp}
\begin{aligned}
&\widehat{\mathbf{x}}| \widetilde{\mathbf{x}}, \widetilde{\mathbf{P}}, \widehat{\mathbf{P}}   \sim \mathcal{N} (\widehat{\boldsymbol{\mu}}_x^{\textnormal{(CGP)}}, \widehat{\textbf{C}}_x^{\textnormal{(CGP)}}), \text{ where} \\
&\widehat{\boldsymbol{\mu}}_x^{\textnormal{(CGP)}}  = \mathbf{\Phi}^{\dagger} \widetilde{\mathbf{\Phi}}^{-1} \widetilde{\mathbf{x}}, \, \, \text{ and } \, \, 
\widehat{\textbf{C}}_x^{\textnormal{(CGP)}} = \widehat{\mathbf{\Phi}} - \mathbf{\Phi}^{\dagger} \widetilde{\mathbf{\Phi}}^{-1} (\mathbf{\Phi}^{\dagger})^T.
\end{aligned}
\end{equation}
Eq. (\ref{eq_pred_sgp}) gives us the predicted mean $\widehat{\boldsymbol{\mu}}_x^{\textnormal{(CGP)}}$ and the associated covariance $\widehat{\textbf{C}}_x^{\textnormal{(CGP)}}$ of the test $x-$coordinate vector $\widehat{\mathbf{x}}$ when the CGP method is employed. The predictive distribution of the $x-$coordinate $[\widehat{\mathbf{x}}]_l$ of any particular test user $l$ can be obtained, through marginalization of the joint predictive distribution of $\widehat{\mathbf{x}}| \widetilde{\mathbf{x}}, \widetilde{\mathbf{P}}, \widehat{\mathbf{P}}$ given by (\ref{eq_pred_sgp}), as
\begin{align} \label{eq_pred_sgp_user}
& \text{$[\widehat{\mathbf{x}}]_l$} | \widetilde{\mathbf{x}}, \widetilde{\mathbf{P}}, \widehat{\mathbf{p}}_l   \sim \mathcal{N} ([\widehat{\boldsymbol{\mu}}_x^{\textnormal{(CGP)}}]_l, [\widehat{\textbf{C}}_x^{\textnormal{(CGP)}}]_{ll}), \text{ where } \nonumber \\
& [\widehat{\boldsymbol{\mu}}_x^{\textnormal{(CGP)}}]_l  =  [\mathbf{\Phi}^{\dagger} \widetilde{\mathbf{\Phi}}^{-1} \widetilde{\mathbf{x}}]_l, \nonumber \\
& \overset{\mathrm{(a)}}{=} \sum_{i=1}^{\widetilde{L}} \phi(\widehat{\mathbf{p}}_l, \widetilde{\mathbf{p}}_i) [\boldsymbol{\psi}]_i , \,\text{ (defined $\boldsymbol{\psi} =  \widetilde{\mathbf{\Phi}}^{-1} \widetilde{\mathbf{x}}$), and} \nonumber \\
& [\widehat{\textbf{C}}_x^{\textnormal{(CGP)}}]_{ll} = [\widehat{\mathbf{\Phi}} - \mathbf{\Phi}^{\dagger} \widetilde{\mathbf{\Phi}}^{-1} (\mathbf{\Phi}^{\dagger})^T]_{ll}. \nonumber \\
& \overset{\mathrm{(b)}}{=} \phi(\widehat{\mathbf{p}}_l, \widehat{\mathbf{p}}_l) - \sum_{i=1}^{\widetilde{L}} \sum_{j=1}^{\widetilde{L}} \phi(\widehat{\mathbf{p}}_l, \widetilde{\mathbf{p}}_i) [(\widetilde{\mathbf{\Phi}} )^{-1}]_{ij} \phi(\widetilde{\mathbf{p}}_j, \widehat{\mathbf{p}}_l).
\end{align}
\noindent In (\ref{eq_pred_sgp_user}), (a)-(b) are obtained by substituting the covariance matrices defined in (\ref{eq_cov_matrices}). Also, the terms $[\widehat{\boldsymbol{\mu}}_x^{\textnormal{(CGP)}}]_l$ and $[\widehat{\textbf{C}}_x^{\textnormal{(CGP)}}]_{ll}$ refer to the predicted mean and variance of the $x-$coordinate $[\widehat{\mathbf{x}}]_l$ of any test user $l$. Since the mean of a Gaussian distribution is also its mode, the predicted mean $[\widehat{\boldsymbol{\mu}}_x^{\textnormal{(CGP)}}]_l$ gives us the maximum-a-posteriori (MAP) estimate of $[\widehat{\mathbf{x}}]_l$. Also, the predictive variance $[\widehat{\textbf{C}}_x^{\textnormal{(CGP)}}]_{ll}$ gives us the $2\sigma$ error-bars $\pm 2 \sqrt {[\widehat{\textbf{C}}_x^{\textnormal{(CGP)}}]_{ll}}$ on choosing $[\widehat{\boldsymbol{\mu}}_x^{\textnormal{(CGP)}}]_l$ as the estimate of $[\widehat{\mathbf{x}}]_l$.  

The CGP method detailed above serves as a baseline method to predict the locations of test users from their noisy RSS vectors. As may be observed from (\ref{eq_joint_dist}), the CGP method naively treats the noisy test RSS data as noise-free, and is therefore, only able to provide location estimates with unrealistically small $2\sigma$ error-bars, even if the predicted locations are erroneous. We will validate this through simulation studies in Section \ref{sec_sim}. This shortcoming can be overcome by the NaGP method discussed in the next section because it accounts for the noisy nature of the test RSS vectors.

\section{Location Prediction With Numerical Approximation GP Method} \label{sec_nagp_gagp}
We now consider the numerical approximation GP method (NaGP), which is a moment matching-based GP method, to estimate the test user locations. This method exploits the stochastic nature of the noisy test RSS vectors to provide more realistic $2\sigma$ error-bars on the estimated locations than the CGP method. Specifically, for each test user $l$, the NaGP method (i) derives the true predictive distribution $p([\widehat{\mathbf{x}}]_l | \widetilde{\mathbf{x}}, \widetilde{\mathbf{P}}, \widehat{\mathbf{p}}_l)$ by taking the input test RSS distribution into account, and then (ii) employs moment matching to numerically approximate the true predictive distribution as Gaussian. Let us first derive the true predictive distribution. 

 \color{black}
We observe from (\ref{eq_rss_dB}) that any noisy RSS value $p_{mk}^{\textnormal{dB}}$ recorded at the RRH $m$ is the sum of a noise-free component, i.e., $p_0^{\text{dB}} - 10 \eta \log_{10} (d_{mk}) $, and a shadowing noise component, i.e., $z_{mk}$. This allows us to express any noisy test RSS vector $\widehat{\mathbf{p}}_l$ as 
\begin{equation} \label{eq_uncertain_rss}
\widehat{\mathbf{p}}_l  = \widehat{\mathbf{p}}_l^* + \widehat{\mathbf{z}}_l, \text{ such that } \widehat{\mathbf{z}}_l \sim \mathcal{N} (\mathbf{0}, \widehat{\boldsymbol{\Sigma}}_l),
\end{equation}
where $\widehat{\mathbf{p}}_l^*$ is the noise-free component in $\widehat{\mathbf{p}}_l$ and $\widehat{\mathbf{z}}_l$ is the shadowing noise with covariance $\widehat{\boldsymbol{\Sigma}}_l$. For simplicity, we assume that $\widehat{\boldsymbol{\Sigma}}_l$ is a diagonal matrix, in other words, we assume that the $M$ uplink channels of the test user $l$ experience mutually independent shadowing. We also assume that the diagonal elements of $\widehat{\boldsymbol{\Sigma}}_l$, which represent the shadowing variances of the $M$ uplink channels of the test user $l$, are already known to the CU. We then know from (\ref{eq_uncertain_rss}) that $\widehat{\mathbf{p}}_l^*$ is conditionally distributed as
\begin{equation} \label{eq_cond_rss}
\widehat{\mathbf{p}}_l^*|\widehat{\mathbf{p}}_l, \widehat{\boldsymbol{\Sigma}}_l \sim \mathcal{N} (\widehat{\mathbf{p}}_l, \widehat{\boldsymbol{\Sigma}}_l ).
\end{equation}
We can now treat $\widehat{\mathbf{p}}_l^*$ as a hidden variable and use (\ref{eq_pred_sgp_user}) to obtain an estimate of $[\widehat{\mathbf{x}}]_l$ in terms of $\widehat{\mathbf{p}}_l^*$. Followed by this, we can use (\ref{eq_cond_rss}) to integrate out the hidden variable $\widehat{\mathbf{p}}_l^*$ and obtain the true predictive distribution of $[\widehat{\mathbf{x}}]_l$ in terms of $\widehat{\mathbf{p}}_l$ as follows\footnotemark{}
\footnotetext{For notational ease, all integrals henceforth are written as indefinite integrals, but in reality, they are definite integrals over appropriate sets.}
\begin{equation} \label{eq_integral_pred}
p ([\widehat{\mathbf{x}}]_l | \widetilde{\mathbf{x}}, \widetilde{\mathbf{P}}, \widehat{\mathbf{p}}_l) = \int p([\widehat{\mathbf{x}}]_l | \widetilde{\mathbf{x}}, \widetilde{\mathbf{P}}, \widehat{\mathbf{p}}_l^*) p(\widehat{\mathbf{p}}_l^*| \widehat{\mathbf{p}}_l, {\widehat{\boldsymbol{\Sigma}}_l}) d \widehat{\mathbf{p}}_l^*,
\end{equation}
where $p([\widehat{\mathbf{x}}]_l | \widetilde{\mathbf{x}}, \widetilde{\mathbf{P}}, \widehat{\mathbf{p}}_l^*)$ is obtained from (\ref{eq_pred_sgp_user}) and $p(\widehat{\mathbf{p}}_n^*| \widehat{\mathbf{p}}_n, {\widehat{\boldsymbol{\Sigma}}_n})$ from (\ref{eq_cond_rss}), respectively. The predictive distribution $p ([\widehat{\mathbf{x}}]_l | \widetilde{\mathbf{x}}, \widetilde{\mathbf{P}}, \widehat{\mathbf{p}}_l)$ in (\ref{eq_integral_pred}) is non-Gaussian and cannot be obtained in closed-form because the integral on the right hand side is intractable.
As a consequence, we can only obtain an approximation to the true predictive distribution $p ([\widehat{\mathbf{x}}]_l | \widetilde{\mathbf{x}}, \widetilde{\mathbf{P}}, \widehat{\mathbf{p}}_l)$, using either numerical or analytical approximation procedures. 

The NaGP method takes a numerical approach and approximates the true predictive distribution $p ([\widehat{\mathbf{x}}]_l  | \widetilde{\mathbf{x}}, \widetilde{\mathbf{P}}, \widehat{\mathbf{p}}_l)$ in (\ref{eq_integral_pred}) using Markov-Chain Monte-Carlo sampling\cite{monte_carlo} as follows. We draw $S$ independent and identically distributed (i.i.d) samples $\widehat{\mathbf{p}}_l^{*} (s)$, $1 \leq s \leq S$, from $\widehat{\mathbf{p}}_l^{*}|\widehat{\mathbf{p}}_l, {\widehat{\boldsymbol{\Sigma}}_l} \sim \mathcal{N}( \widehat{\mathbf{p}}_l, {\widehat{\boldsymbol{\Sigma}}_l})$ and approximate the integral in (\ref{eq_integral_pred}) as
\begin{equation} \label{eq_mcgp}
\begin{aligned}
p ([\widehat{\mathbf{x}}]_l  | \widetilde{\mathbf{x}}, \widetilde{\mathbf{P}}, \widehat{\mathbf{p}}_l) & \overset{\mathrm{(a)}}{\approx}  \sum_{s=1}^S  \frac{1}{S} p([\widehat{\mathbf{x}}]_l | \widetilde{\mathbf{x}}, \widetilde{\mathbf{P}}, \widehat{\mathbf{p}}_l^{*} (s)) \\
&  \overset{\mathrm{(b)}}{=}  \sum_{s=1}^S  \frac{1}{S} \mathcal{N} ([\widehat{\mathbf{x}}]_l; [\widehat{\boldsymbol{\mu}}_{x}^{\textnormal{CGP}} (s)]_l, [\widehat{\mathbf{C}}_{x}^{\textnormal{CGP}} (s)]_{ll} ), \\
\end{aligned}
\end{equation}
\noindent where (a) follows from the Monte-Carlo approximation procedure \cite{monte_carlo}, and (b) from (\ref{eq_pred_sgp_user}), with the $[\widehat{\boldsymbol{\mu}}_{x}^{\textnormal{CGP}} (s) ]_l$ and $[\widehat{\mathbf{C}}_{x}^{\textnormal{CGP}} (s) ]_{ll}$ being the same as $[\widehat{\boldsymbol{\mu}}_x^{\textnormal{CGP}}]_l$ and $[\widehat{\mathbf{C}}_x^{\textnormal{CGP}}]_{ll}$ respectively, but with the test RSS vector $\widehat{\mathbf{p}}_l$ replaced by the Monte-Carlo sample $\widehat{\mathbf{p}}_l^{*} (s)$. Since the right hand side of (\ref{eq_mcgp}) is a mixture of $S$ Gaussians with identical weights, we know from  \cite{gauss_mixture} (eq. (14.10)-(14.11)) that we can approximate the left hand side as a Gaussian distribution with the same first and second order moments, as given below
\begin{equation} \label{eq_mcgp_approx}
\begin{aligned}
& p ([\widehat{\mathbf{x}}]_l  | \widetilde{\mathbf{x}}, \widetilde{\mathbf{P}}, \widehat{\mathbf{p}}_l)  \approx \mathcal{N} ([\widehat{\mathbf{x}}]_l; [\widehat{\boldsymbol{\mu}}_x^{\textnormal{(NaGP)}}]_l, [\widehat{\mathbf{C}}_x^{\textnormal{(NaGP)}}]_{ll}), \text{where, }  \\
& \text{$[\widehat{\boldsymbol{\mu}}_x^{\textnormal{(NaGP)}}]_l$} = \frac{1}{S} \sum_{s=1}^S  [\widehat{\boldsymbol{\mu}}_{x}^{\textnormal{CGP}}(s) ]_l, \\
& [\widehat{\mathbf{C}}_x^{\textnormal{(NaGP)}}]_{ll}  = \frac{1}{S} \sum_{s=1}^S ([\widehat{\boldsymbol{\mu}}_{x}^{\textnormal{CGP}} (s) ]_l - [\widehat{\boldsymbol{\mu}}_x^{\textnormal{(NaGP)}}]_l)^2 \\
& \quad \quad \quad \quad \quad + \frac{1}{S} \sum_{s=1}^S [\widehat{\mathbf{C}}_{x}^{\textnormal{CGP}} (s) ]_{ll}, \forall l = 1, \dots, \widehat{L}.
\end{aligned}
\end{equation}
In (\ref{eq_mcgp_approx}), $[\widehat{\boldsymbol{\mu}}_x^{\textnormal{(NaGP)}}]_l$ refers to the estimate of the test $x-$coordinate $[\widehat{\mathbf{x}}]_l$ from NaGP and $[\widehat{\textbf{C}}_x^{\textnormal{(NaGP)}}]_{ll}$ refers to the associated variance.
By increasing $S$, we can increase the accuracy of the $[\widehat{\boldsymbol{\mu}}_x^{\textnormal{(NaGP)}}]_l$ and $[\widehat{\textbf{C}}_x^{\textnormal{(NaGP)}}]_{ll}$ values because the numerical approximation procedure in (\ref{eq_mcgp}) becomes tighter with increasing $S$ \cite{monte_carlo}.

\color{black}

\begin{remark}
We may note from (\ref{eq_integral_pred}) that, unlike the CGP method which naively treats the noisy test RSS vectors as noise-free, the NaGP method treats the noise-free components in the test RSS vectors as hidden variables and integrates them out using statistical knowledge of the noise present. By doing so, the NaGP method learns from the noise present in the test RSS vectors and incorporates their noise covariance matrices $\{\widehat{\boldsymbol{\Sigma}}_l\}$ into the predicted mean and variance expressions (c.f. (\ref{eq_mcgp_approx})). This learning allows the NaGP method to provide more realistic $2\sigma$ error-bars on the predicted locations than the CGP method. 
\end{remark}

In the next section, we present details on the performance metrics considered and also derive a Cramer-Rao lower bound on the achievable root-mean-squared error performance of the two GP methods under study.

\section{Performance Metrics and Cramer-Rao Lower Bound} \label{sec_perf}
We measure prediction performance in terms of (i) the root-mean-squared prediction error (RMSE) and (ii) the log-predictive density (LPD), defined as

\begin{align} \label{eq_rmse_lpd}
\text{RMSE} &= \sqrt{\frac{\sum\limits_{l=1}^{\widehat{L}}([\widehat{\mathbf{x}}]_l - [\widehat{\boldsymbol{\mu}}_{x}^{(.)}]_l)^2 + ([\widehat{\mathbf{y}}]_l - [\widehat{\boldsymbol{\mu}}_{y}^{(.)}]_l)^2}{\widehat{L}}}, \text{ and} \nonumber \\
\text{LPD} &= \frac{1}{{\widehat{L}}} (\log (p(\widehat{\mathbf{x}} |\widetilde{\mathbf{x}}, \widetilde{\mathbf{P}}, \widehat{\mathbf{P}})) + \log (p(\widehat{\mathbf{y}} |\widetilde{\mathbf{y}}, \widetilde{\mathbf{P}}, \widehat{\mathbf{P}}))) , \nonumber \\
&= - \log (2\pi) - \frac{1}{2 {\widehat{L}}} \sum_{l=1}^{\widehat{L}} \Bigg \{ \log ([\widehat{\mathbf{C}}_x^{(.)}]_{ll})  + \log ([\widehat{\mathbf{C}}_y^{(.)}]_{ll})  +\nonumber \\
& \frac{([\widehat{\mathbf{x}}]_l - [\widehat{\boldsymbol{\mu}}_{x}^{(.)}]_l)^2}{[\widehat{\mathbf{C}}_x^{(.)}]_{ll}} +  \frac{([\widehat{\mathbf{y}}]_l - [\widehat{\boldsymbol{\mu}}_{y}^{(.)}]_l)^2}{[\widehat{\mathbf{C}}_y^{(.)}]_{ll}} \Bigg \}, 
\end{align} 
where $[\widehat{\mathbf{x}}]_l$ and $[\widehat{\mathbf{y}}]_l$ are the actual $x$ and $y$ coordinates of the test user $n$, $[\widehat{\boldsymbol{\mu}}_{x}^{(.)}]_l$ and $[\widehat{\boldsymbol{\mu}}_{y}^{(.)}]_l$ are the estimates of $[\widehat{\mathbf{x}}]_l$ and $[\widehat{\mathbf{y}}]_l$ given by the chosen GP method, and $[\widehat{\mathbf{C}}_{x}^{(.)}]_{ll}$ and $[\widehat{\mathbf{C}}_{y}^{(.)}]_{ll}$ are the variances associated with the estimates $[\widehat{\boldsymbol{\mu}}_{x}^{(.)}]_l$ and $[\widehat{\boldsymbol{\mu}}_{y}^{(.)}]_l$, respectively. For example, if we choose the CGP method, $[\widehat{\boldsymbol{\mu}}_{x}^{(.)}]_l = [\widehat{\boldsymbol{\mu}}_{x}^{(\textnormal{CGP})}]_l$, $[\widehat{\boldsymbol{\mu}}_{y}^{(.)}]_l = [\widehat{\boldsymbol{\mu}}_{y}^{(\textnormal{CGP})}]_l$, $[\widehat{\mathbf{C}}_{x}^{(.)}]_{ll} = [\widehat{\mathbf{C}}_{x}^{(\textnormal{CGP})}]_{ll}$ and $[\widehat{\mathbf{C}}_{y}^{(.)}]_{ll} = [\widehat{\mathbf{C}}_{y}^{(\textnormal{CGP})}]_{ll}$. The RMSE metric only takes the estimates $[\widehat{\boldsymbol{\mu}}_{x}^{(.)}]_l$ and $[\widehat{\boldsymbol{\mu}}_{y}^{(.)}]_l$  into account and ignores the uncertainties $[\widehat{\mathbf{C}}_{x}^{(.)}]_{ll}$ and $[\widehat{\mathbf{C}}_{y}^{(.)}]_{ll}$ around them. In contrast, the LPD metric takes the entire predictive distribution into account. Observe from (\ref{eq_rmse_lpd}) that the LPD metric penalizes overconfident location estimates by assigning larger weights to the prediction errors $([\widehat{\mathbf{x}}]_l - [\widehat{\boldsymbol{\mu}}_{x}^{(.)}]_l)$ and $([\widehat{\mathbf{y}}]_l - [\widehat{\boldsymbol{\mu}}_{y}^{(.)}]_l)$ when the associated uncertainties $[\widehat{\mathbf{C}}_{x}^{(.)}]_{ll}$ and $[\widehat{\mathbf{C}}_{y}^{(.)}]_{ll}$ are small. Lower RMSE values and higher LPD values indicate better prediction performance.

\subsection{Cramer-Rao Lower Bound on the RMSE Performance}
 To evaluate the RMSE performance of the presented GP methods, we need a theoretical lower bound on the achievable RMSE performance. Towards this, we derive a Bayesian Cramer-Rao lower bound that reflects the best possible RMSE performance of any unbiased estimator of the test user's location coordinates. 
 
 The location prediction problem under study can be viewed as an estimation problem in which we wish to estimate the test user $x-$coordinate vector $\widehat{\mathbf{x}}$ from the training $x-$coordinate measurements $\widetilde{\mathbf{x}}$, given the training RSS data $\widetilde{\mathbf{P}}$, the test RSS data $\widehat{\mathbf{P}}$, and the free parameter vector $\boldsymbol{\theta}$ (available upon training the GP model). 
 Therefore, for a chosen GP method, the Bayesian Cramer Rao lower bound (BCRLB) on the expected squared-error matrix for the test users' $x-$coordinates is given by \cite{trees_book}
 \begin{equation} \label{eq_bcrb}
 \begin{aligned}
 & \mathbb{E} ((\widehat{\mathbf{x}} - \widehat{\boldsymbol{\mu}}_x^{(.)}) (\widehat{\mathbf{x}} - \widehat{\boldsymbol{\mu}}_x^{(.)})^T) \succeq \textnormal{BCRLB}_{x}, \text{where,}\\
& \textnormal{BCRLB}_{x} =  - (\mathbb{E} (\nabla_{\widehat{\mathbf{x}} } [\nabla_{\widehat{\mathbf{x}} }\log (p(\widetilde{\mathbf{x}}, \widehat{\mathbf{x}} | \widetilde{\mathbf{P}}, \widehat{\mathbf{P}}, \boldsymbol{\theta}))]^T))^{-1}. 
 \end{aligned}
 \end{equation}
In (\ref{eq_bcrb}), $\widehat{\boldsymbol{\mu}}_x^{(.)}$ is the estimate of $\widehat{\mathbf{x}}$ provided by the chosen GP method and $\textnormal{BCRLB}_{x}$ is the associated BCRLB. The expectation $\mathbb{E} (.)$ is with respect to the training users' $x-$coordinate vector $\widetilde{\mathbf{x}}$ and the test users' $x-$coordinate vector $\widehat{\mathbf{x}}$. The term $\mathbb{E} (\nabla_{\widehat{\mathbf{x}} } (\nabla_{\widehat{\mathbf{x}} }(\log p(\widetilde{\mathbf{x}},\widehat{\mathbf{x}}  |  \widetilde{\mathbf{P}}, \widehat{\mathbf{P}},\boldsymbol{\theta}))))$ on the right hand side of (\ref{eq_bcrb}) is the Bayesian Information Matrix (BIM) on $\widehat{\mathbf{x}} $ \cite{trees_book}, which we simplify as follows
 \begin{align} \label{eq_deriv_bcrb_rhs}
 & \mathbb{E} (\nabla_{\widehat{\mathbf{x}} } [\nabla_{\widehat{\mathbf{x}} }(\log (p(\widetilde{\mathbf{x}}, \widehat{\mathbf{x}}  |  \widetilde{\mathbf{P}}, \widehat{\mathbf{P}}, \boldsymbol{\theta})))]^T) \nonumber \\
 & \overset{\mathrm{(a)}}{=} \mathbb{E} (\nabla_{\widehat{\mathbf{x}} } [\nabla_{\widehat{\mathbf{x}} } \log (p( \widehat{\mathbf{x}}  | \widetilde{\mathbf{x}},  \widetilde{\mathbf{P}}, \widehat{\mathbf{P}}, \boldsymbol{\theta})) + \nabla_{\widehat{\mathbf{x}} } \log (p(\widetilde{\mathbf{x}} |  \widetilde{\mathbf{P}}, \widehat{\mathbf{P}}, \boldsymbol{\theta} ))]^T) \nonumber \\
 & \overset{\mathrm{(b)}}{=} \mathbb{E} (\nabla_{\widehat{\mathbf{x}} } [\nabla_{\widehat{\mathbf{x}} } (-\frac{\widehat{L}}{2} \log  (2 \pi) - \frac{1}{2} (|\widehat{\mathbf{C}}_x^{(.)}|) - \frac{1}{2} (\widehat{\mathbf{x}} - \widehat{\boldsymbol{\mu}}_x^{(.)})^T\nonumber \\
 & \quad (\widehat{\mathbf{C}}_x^{(.)})^{-1} (\widehat{\mathbf{x}} - \widehat{\boldsymbol{\mu}}_x^{(.)})) + \nabla_{\widehat{\mathbf{x}} } ( -\frac{\widetilde{L}}{2} \log (2\pi) - \frac{1}{2} \log |\widetilde{\mathbf{\Phi}}| - \nonumber \\ 
 &\quad \frac{1}{2} \widetilde{\mathbf{x}}^T \widetilde{\mathbf{\Phi}}^{-1}  \widetilde{\mathbf{x}} ) ]^T)  \nonumber \\
 & \overset{\mathrm{(c)}}{=} - \mathbb{E} (\nabla_{\widehat{\mathbf{x}} } [ \nabla_{\widehat{\mathbf{x}} } ( \frac{1}{2} (\widehat{\mathbf{x}} - \widehat{\boldsymbol{\mu}}_x^{(.)})^T(\widehat{\mathbf{C}}_x^{(.)})^{-1} (\widehat{\mathbf{x}} - \widehat{\boldsymbol{\mu}}_x^{(.)})) ]^T) \nonumber \\
 & \overset{\mathrm{(d)}}{=} - \mathbb{E} ((\widehat{\mathbf{C}}_x^{(.)})^{-1}) \nonumber \\
 & \overset{\mathrm{(e)}}{=} - (\widehat{\mathbf{C}}_x^{(.)})^{-1}, 
 \end{align}
 \noindent where (a) follows from Bayes' rule, (b) from substituting $\, p( \widehat{\mathbf{x}}  | \widetilde{\mathbf{x}}, \widetilde{\mathbf{P}}, \widehat{\mathbf{P}}, \boldsymbol{\theta})$ from the chosen GP method (c.f. \textbf{Remark \ref{remark_bcrb_choice}} below) and $p(\widetilde{\mathbf{x}} |\widetilde{\mathbf{P}},\widehat{\mathbf{P}}, \boldsymbol{\theta} ) = p(\widetilde{\mathbf{x}} |\widetilde{\mathbf{P}}, \boldsymbol{\theta} )$ from (\ref{eq_trainingdist}), (c) from setting the gradient $\nabla_{\widehat{\mathbf{x}} } (.)$ of all the terms which are constant with respect to (w.r.t) $\widehat{\mathbf{x}} $ to zero, (d) from evaluating the gradient twice w.r.t $\widehat{\mathbf{x}} $, and (e) from observing that the elements of $(\widehat{\mathbf{C}}_x^{(.)})^{-1}$ are independent of both $\widetilde{\mathbf{x}}$ and $\widehat{\mathbf{x}} $ (please see the expressions for $\widehat{\mathbf{C}}_x^{(\textnormal{CGP})}$ in (\ref{eq_pred_sgp_user}) and $\widehat{\mathbf{C}}_x^{(\textnormal{NaGP})}$ in (\ref{eq_mcgp_approx})). 
 \noindent Substituting (\ref{eq_deriv_bcrb_rhs}) into (\ref{eq_bcrb}), we have
 \begin{equation} \label{eq_rmse_bcrb1}
  \textnormal{BCRLB}_x = \widehat{\mathbf{C}}_x^{(.)}.
 \end{equation}
 Similarly, we can obtain the BCRLB for the expected squared-error matrix  of the test users' $y-$coordinates as
 \begin{equation}\label{eq_rmse_bcrb2}
 \textnormal{BCRLB}_y = \widehat{\mathbf{C}}_y^{(.)}.
 \end{equation}
 From (\ref{eq_rmse_bcrb1}) and (\ref{eq_rmse_bcrb2}), we can obtain a Bayesian Cramer-Rao lower bound on the RMSE for predicting the test user locations as follows
 \begin{equation} \label{eq_lower_bound}
 \begin{aligned}
 \textnormal{BCRLB}^\textnormal{(RMSE)} & =\sqrt{ \frac{1}{\widehat{L}} \Tr (\textnormal{BCRLB}_x + \textnormal{BCRLB}_y)} \\
 & = \sqrt{ \frac{1}{\widehat{L}} \Tr (\widehat{\mathbf{C}}_x^{(.)} + \widehat{\mathbf{C}}_y^{(.)})},
 \end{aligned}
 \end{equation}
 where $\widehat{L}$ is the number of test users. Eq. (\ref{eq_lower_bound}) shows that the $\textnormal{BCRLB}^\textnormal{(RMSE)}$ for any chosen GP method can be obtained from its predictive covariances $\widehat{\mathbf{C}}_x^{(.)}$ and $\widehat{\mathbf{C}}_y^{(.)}$ through simple linear algebraic operations. To obtain a valid $\textnormal{BCRLB}^\textnormal{(RMSE)}$, we must therefore ensure that the $\widehat{\mathbf{C}}_x^{(.)}$ and $\widehat{\mathbf{C}}_y^{(.)}$ values are accurate. \textbf{Remark \ref{remark_bcrb_choice}} below summarizes our approach to obtain accurate $\widehat{\mathbf{C}}_x^{(.)}$ and $\widehat{\mathbf{C}}_y^{(.)}$ values for calculating the $\textnormal{BCRLB}^\textnormal{(RMSE)}$ for both the presented GP methods.
 
 \begin{remark} \label{remark_bcrb_choice}
	As explained in Section \ref{sec_nagp_gagp}, the true predictive distribution $p( \widehat{\mathbf{x}}  | \widetilde{\mathbf{x}}, \widetilde{\mathbf{P}}, \widehat{\mathbf{P}}, \boldsymbol{\theta})$ cannot be obtained in exact form. Nevertheless, the NaGP method in Section \ref{sec_nagp_gagp} gives us a numerical approximation for $p( \widehat{\mathbf{x}}  | \widetilde{\mathbf{x}}, \widetilde{\mathbf{P}}, \widehat{\mathbf{P}}, \boldsymbol{\theta})$ (c.f. (\ref{eq_mcgp})). Therefore, to calculate the $\textnormal{BCRLB}^\textnormal{(RMSE)}$ for both the CGP and NaGP methods using (\ref{eq_lower_bound}), we advocate the use of $\widehat{\mathbf{C}}_x^{(\textnormal{NaGP})}$ (and $\widehat{\mathbf{C}}_y^{(\textnormal{NaGP})}$) obtained from (\ref{eq_mcgp_approx}) as the  $\widehat{\mathbf{C}}_x^{(.)}$ (and $\widehat{\mathbf{C}}_y^{(.)}$).
 \end{remark}

 \begin{figure}
 \centering
 		\includegraphics[scale=0.63] {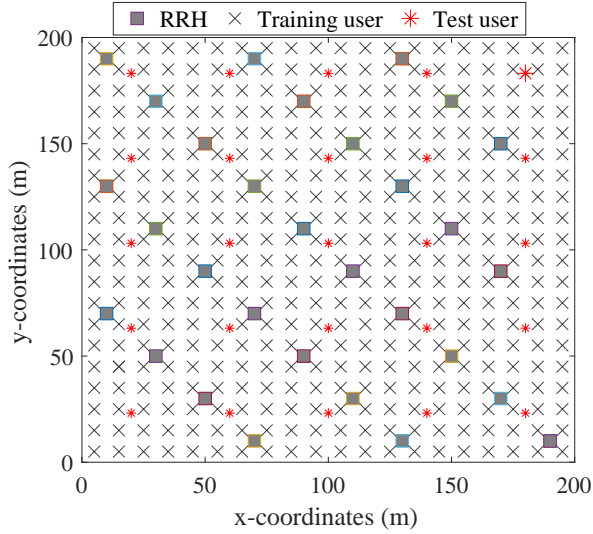}
 		\caption{Simulation setup with $M = 30$ RRH antennas, $\widetilde{L} = 400$ training user locations, and $\widehat{L} = 25$ test user locations.}
 		\label{fig_sim_setup}
 \end{figure}

\begin {table} 
\caption{ Parameters for simulation studies \label{table_pathloss} }
\fontsize{12}{12}
\centering 
\begin{tabular}{| l | l | l |}
	\hline
{System Parameters} & {Value} \\ \hline \hline
 Path-loss parameters & $d_0 = 10$m, \\
(3GPP UMi \cite{3gpp})& $l_0 = -47.5$dB, \\ 
&$\eta = \small {\begin{cases}
	0 \text{ if } d_{mk} < 10\text{m}, \\
	2 \text{ if } 10\text{m} \leq d_{mk} \leq 45\text{m}, \\
	6.7 \text{ if otherwise.} 
	\end{cases}}$ \\ \hline
UE transmit power & $21$dBm ($125$mW) \\ \hline
Noise power& -107.5 dBm \\ \hline
Receiver sensitivity& -106.5 dBm \\ \hline
\end{tabular}
\end {table}
\color{black}

\section{Numerical Studies and Discussions} \label{sec_sim}
We now present numerical examples to evaluate the RMSE and LPD performance of the two GP methods under study, when the shadowing variance $\sigma_z^2$ in the test RSS and the number of RRH antennas $M$ are varied. 

\subsubsection{Parameters and Setup} We consider the example massive MIMO setup shown in Fig. \ref{fig_sim_setup} with $M = 30$ RRH antennas and $\widetilde{L} = 400$ training user locations distributed uniformly over a service area of $200$m $\times$ $200$m. \color{black} All the training user locations are assumed to be available with a measurement error variance ($\sigma_{er}^2$) of $1$dB. The goal is to predict the locations of $\widehat{L} = 25$ test users which are distributed uniformly within the service area. For the training phase, we generate a noise-free training RSS matrix $\widetilde{\mathbf{P}}$ using (\ref{eq_rss_dB}) with shadowing variance $\sigma_z^2 = 0$ and other parameters as per Table \ref{table_pathloss}. Entries of Table \ref{table_pathloss} are chosen as follows. The path-loss parameters $l_0$, $d_0$, and $\eta$ are chosen as per the 3GPP Ubran Micro model \cite{3gpp}. The user transmit power is chosen as per LTE standards to be $21$dBm \cite{tab1}. Total noise power in the system is set to $-107.5$dBm. The uplink receiver sensitivity, which represents the minimum detection threshold for the receiver to distinguish between the signal strength and the noise power, is set to $-106.5$dBm.

\color{black}
 Once the training RSS data is ready, a GP model is trained by solving the log-likelihood maximization problem in (\ref{eq_theta_opt}) using conjugate gradient (CG) method \cite{optim_book}. Multiple trials are run with random initial values to avoid choosing a bad local optimum. Convergence of the CG method is well-known and is therefore skipped here. The same learned parameter vector $\bar{\boldsymbol {\theta}}$ is reused for evaluating the prediction performance of both the GP methods under study because the training dataset and the training procedure are the same.

We generate $200$ Monte-Carlo test RSS matrices each for shadowing variance $\sigma_z^2 = 1, 2,  \dots, 5$dB, using (\ref{eq_rss_dB}) with parameters chosen as per Table \ref{table_pathloss}. During simulations, any instantaneous test RSS value that is lower than the receiver sensitivity is replaced with the noise power in the system. \color{black} The RMSE and LPD values, averaged over the Monte-Carlo realizations, are reported. For the NaGP method, we set the number of Monte-Carlo samples $S = 10$. The CGP method, which naively treats the noisy test RSS data as noise-free for location prediction (c.f. Section \ref{sec_sgp}), serves as the baseline scheme for our analysis.

\begin{figure}
	\begin{center}
		\includegraphics[scale=0.61] {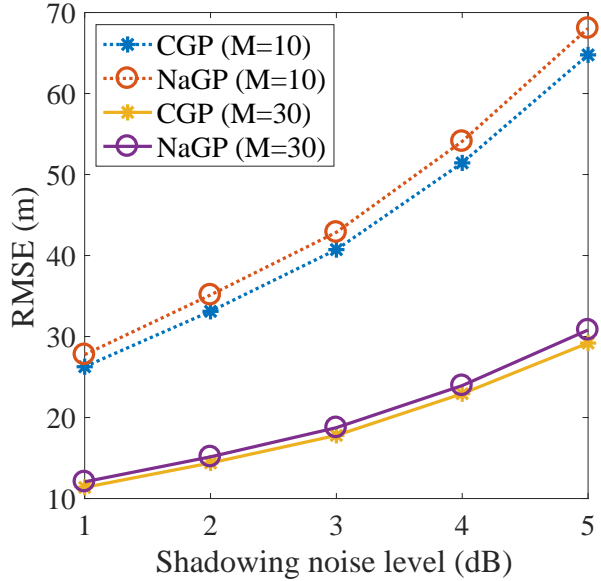}
		\caption{Average RMSE performance of the CGP and NaGP methods for different shadowing noise levels, when $M = 10,$ $30$.}
		\label{figure_rmse}
	\end{center}
\end{figure}
\subsubsection{RMSE Performance} In Fig. \ref{figure_rmse}, we plot the average RMSE achieved by the two GP methods under study, for shadowing variance ($\sigma_z^2$) ranging from $1$dB to $5$dB when the number of RRHs $M = 10$ and $30$. We observe that both the CGP and NaGP methods provide similar RMSE values for different shadowing noise levels. This is because the location estimates from the CGP and the NaGP methods are found to be similar in value. We also observe that the RMSE values increase with the shadowing noise level. This is expected because both the CGP and NaGP methods are trained with noise-free RSS data - they tend to project the noise present in the input RSS onto the output location coordinate space. Lastly, we also observe that the RMSE values are smaller when the number of RRHs $M$ is increased from $10$ to $30$. This clearly reflects the benefit of entering into the massive MIMO regime for RSS-based user positioning. While several studies \cite{marzetta} \cite{ee_paper} have shown that massive MIMO provides large spectral and energy efficiency gains over conventional MIMO systems, we report here that when machine learning methods with noise-free training RSS and noisy test RSS are used to estimate user locations, the massive MIMO technology provides significant gains in the prediction performance over conventional MIMO systems.

\begin{figure*} [t!]
	\captionsetup[subfigure]{justification=centering}
	\begin{subfigure}[t]{.33\textwidth}
		\centering{
			\includegraphics[width=1\linewidth]{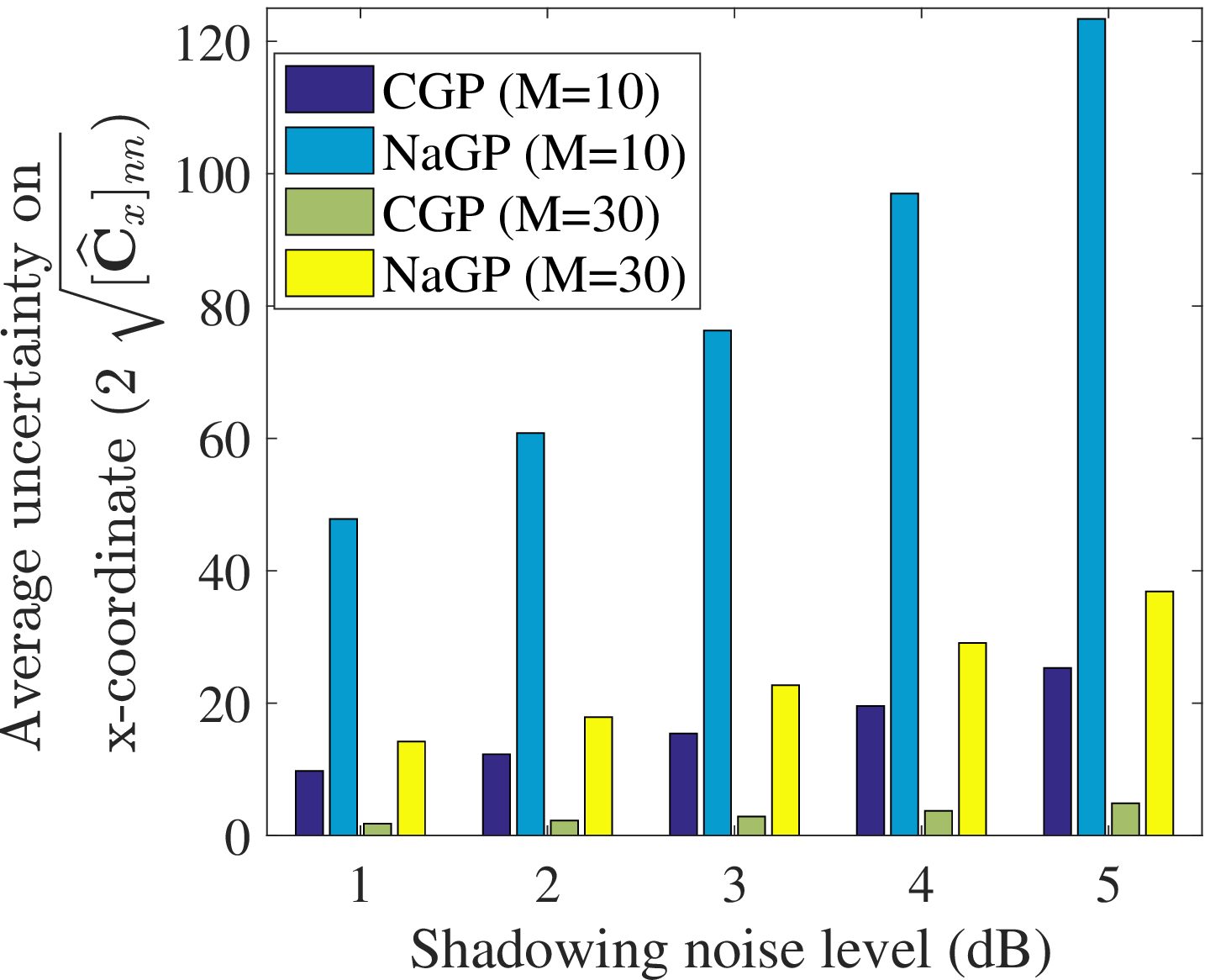}
			\caption{Average $2\sigma$ error-bars on the $x-$coordinates.}
			\label{figure_confx}}
	\end{subfigure}
	\begin{subfigure}[t]{.33\textwidth}
		\centering
		\includegraphics[width=1\linewidth]{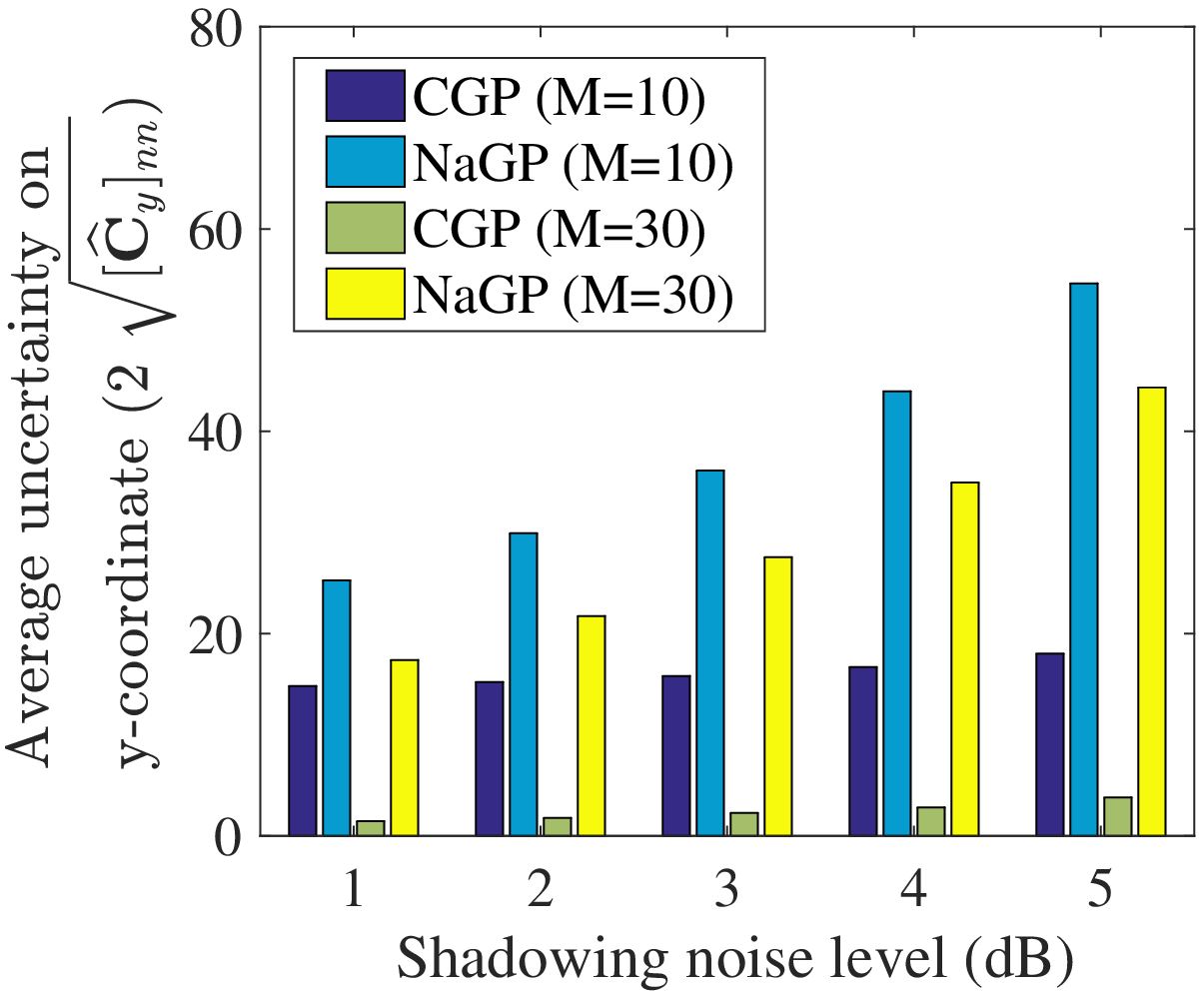}
		\caption{Average $2\sigma$ error-bars on the $y-$coordinates.}
		\label{figure_confy}
	\end{subfigure}%
	\begin{subfigure}[t]{.33\textwidth}
	\centering
	\includegraphics[width=1\linewidth]{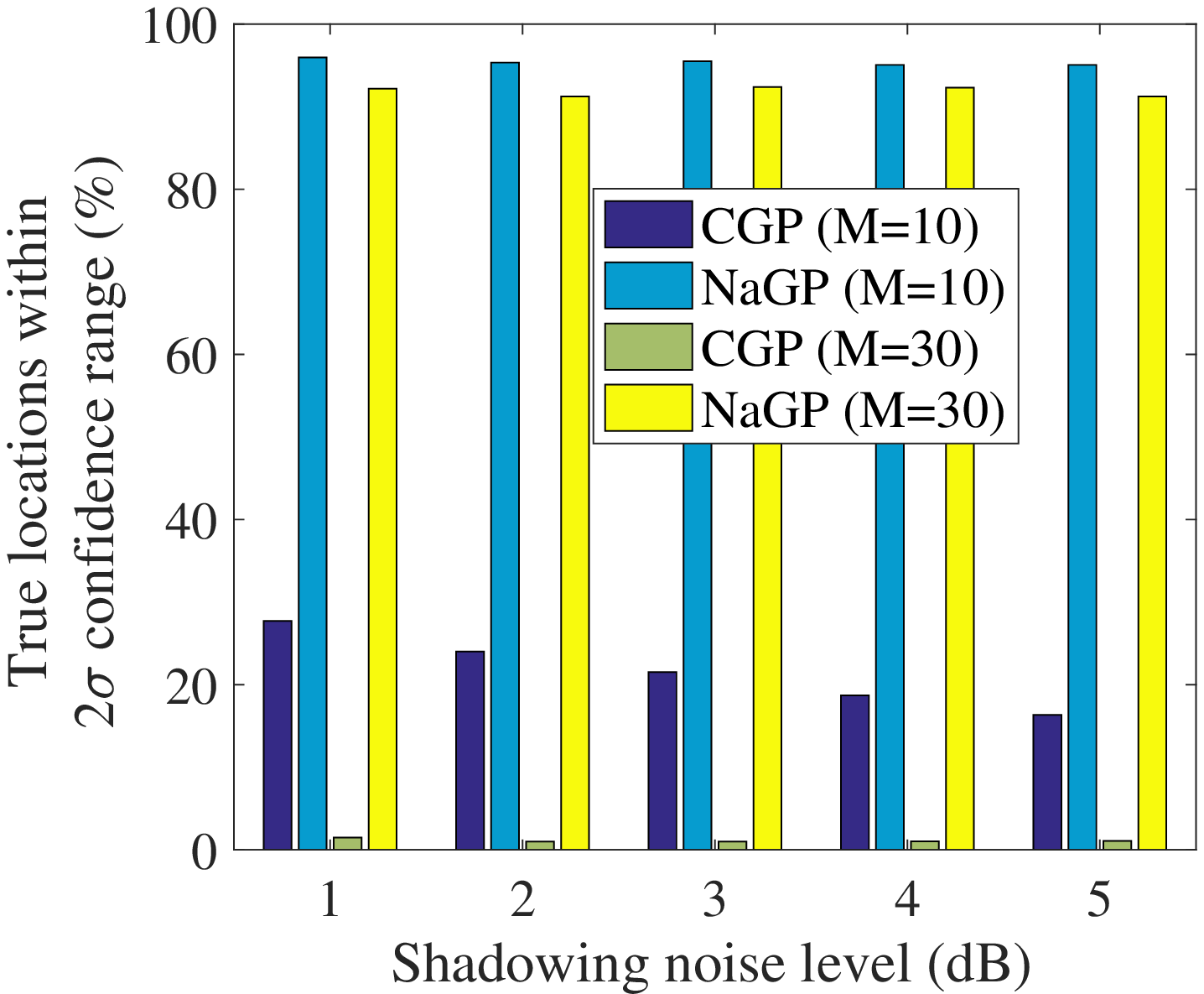}
	\caption{Average true locations within the $2\sigma$ confidence range.}
	\label{figure_true_locs}
\end{subfigure}%
	\caption{Plots of the average $2\sigma$ error-bars on the test users' $x$ and $y$ coordinates and the number of true test user locations within the $2\sigma$ confidence range $([\widehat{\boldsymbol{\mu}}_{x}^{(.)}]_l \pm 2 \sqrt {[\widehat{\textbf{C}}_{x}^{(.)}]_{ll}},  [\widehat{\boldsymbol{\mu}}_{y}^{(.)}]_l \pm 2 \sqrt {[\widehat{\textbf{C}}_{y}^{(.)}]_{ll}})$ of the estimated locations, as provided by the CGP and NaGP methods for different shadowing noise levels, when $M = 10,$ $30$. \label{figure_conf}}
\end{figure*}

\subsubsection{$2\sigma$ Error-Bar Performance} In Fig. \ref{figure_confx} and Fig. \ref{figure_confy}, we plot the average $2\sigma$ error-bars on the test users' $x-$ coordinates and $y-$coordinates, respectively, as given by the CGP and the NaGP methods. In Fig. \ref{figure_true_locs}, we plot the number of true test user locations which are within the $2\sigma$ confidence range $([\widehat{\boldsymbol{\mu}}_{x}^{(.)}]_l \pm 2 \sqrt {[\widehat{\textbf{C}}_{x}^{(.)}]_{ll}},  [\widehat{\boldsymbol{\mu}}_{y}^{(.)}]_l \pm 2 \sqrt {[\widehat{\textbf{C}}_{y}^{(.)}]_{ll}})$ of the estimated locations. We observe that the CGP method provides unrealistically small $2\sigma$ error-bars, which are very low even if the corresponding RMSE values in Fig. \ref{figure_rmse} are high. As a result, less than $30\%$ (for $M=10$) and $10\%$ (for $M = 30$) of the true test user locations are within the $2\sigma$ confidence range provided by the CGP method. In contrast, the NaGP method provides more realistic $2\sigma$ error-bars on the estimated locations. We notice from Fig. \ref{figure_true_locs} that more than $90\%$ of the true test user locations are inside the estimated $2\sigma$ confidence range of the NaGP method for both $M=10$ and $M=30$.

\begin{figure}
	\begin{center}
		\includegraphics[scale=0.61] {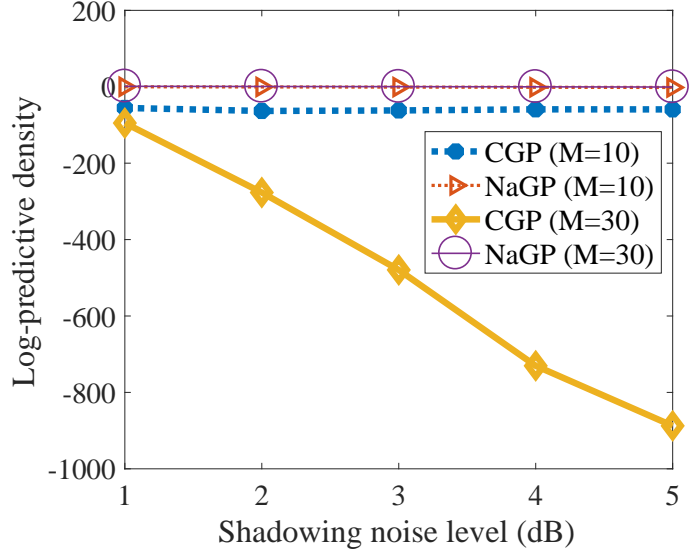}
		\caption{Average LPD performance of the CGP and NaGP methods for different shadowing noise levels, when $M = 10,$ $30$.}
		\label{figure_lpd}
	\end{center}
\end{figure}

\subsubsection{LPD Performance} In Fig. \ref{figure_lpd}, we plot the LPD performance of the CGP and NaGP methods when $M=10$ and $30$. We observe that the CGP method achieves very low LPD values because it provides unrealistically small $[\widehat{\mathbf{C}}_{x}]_{ll}$ and $[\widehat{\mathbf{C}}_{y}]_{ll}$ values (c.f. Fig. \ref{figure_confx} and \ref{figure_confy}), with less than $30\%$ of the true user locations falling inside the $2\sigma$ confidence range of the predicted locations. Note from (\ref{eq_rmse_lpd}) that the LPD metric penalizes such overconfident estimates by assigning large weights to the prediction error. The NaGP method achieves much higher LPD values than the CGP method because it provides realistic $[\widehat{\mathbf{C}}_{x}]_{ll}$ and $[\widehat{\mathbf{C}}_{y}]_{ll}$ values (c.f. Fig. \ref{figure_confx} and \ref{figure_confy}), with more than $90\%$ of the true user locations inside the $2\sigma$ confidence range of the estimated locations (c.f. Fig. \ref{figure_true_locs}).

Taking both the RMSE and LPD plots into perspective, we observe that the NaGP method achieves significantly better LPD performance than the CGP method, while achieving comparable RMSE performance. The superior LPD performance is because the NaGP method learns from the statistical properties of the noise present in the test RSS vectors to provide realistic estimates of the $2\sigma$ error-bars on the predicted locations. 

\begin{figure}
	\begin{center}
		\includegraphics[scale=0.61] {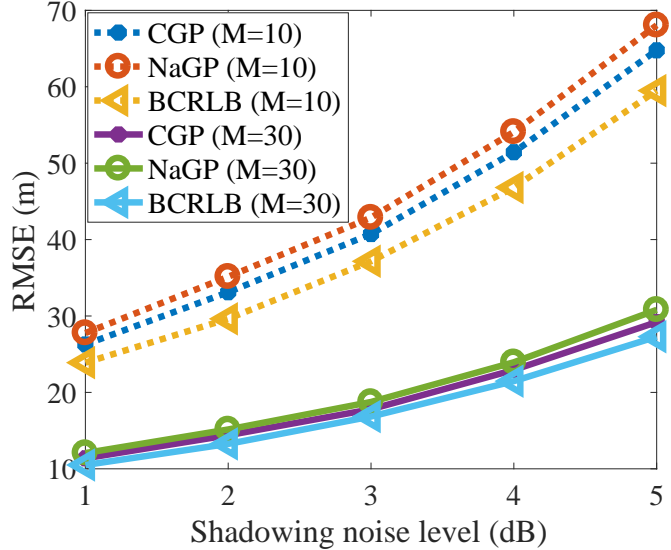}
		\caption{BCRLB on the RMSE performance of the CGP and NaGP methods for different shadowing noise levels, when $M = 10,$ $30$.}
		\label{figure_bcrlb}
	\end{center}
\end{figure}
\subsubsection{Cramer-Rao Lower Bound} In Fig. \ref{figure_bcrlb}, we plot the Bayesian Cramer Rao lower bound on the RMSE performance of the CGP and NaGP methods, with the BCRLB computed using (\ref{eq_lower_bound}) with $\widehat{\mathbf{C}}_x^{(.)} = \widehat{\mathbf{C}}_x^{(\textnormal{NaGP})}$ and $\widehat{\mathbf{C}}_x^{(.)} = \widehat{\mathbf{C}}_y^{(\textnormal{NaGP})}$. We notice that the achieved RMSE performance is very close to the BCRLB, with the bound being tighter for larger $M$. We expect the gap between the achieved RMSE and the BCRLB to be wider for lower number of RRHs $M$ and also for higher shadowing variance $\sigma_{z}^2$ because there is a higher chance of errors caused by thresholding of the test RSS values that are lower than the uplink receiver sensitivity. For large $M$ and/or small $\sigma_{z}^2$, we expect a lesser percentage of RRHs to encounter the test RSS values that are lower than the uplink receiver sensitivity, thus reducing the scope for errors from thresholding.

\begin{figure}
	\begin{center}
		\includegraphics[scale=0.61] {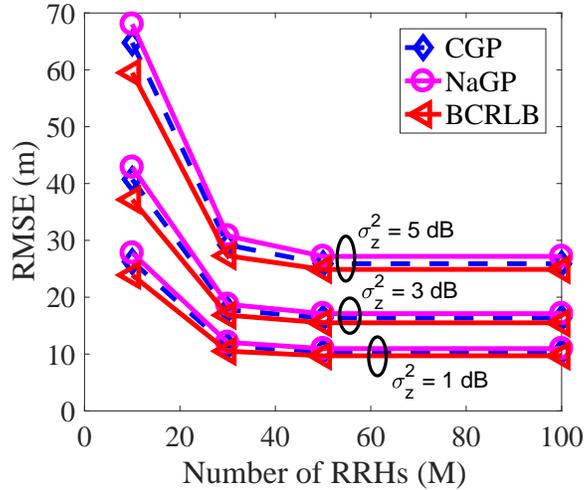}
		\caption{Average RMSE performance of the CGP and NaGP methods and the corresponding BCRLB for different number of RRHs.}
		\label{figure_rmseM}
	\end{center}
\end{figure}
 
\subsubsection{Impact of the number of RRHs} In Fig. \ref{figure_rmseM}, we plot the average RMSE performance of the CGP and NaGP methods, along with their BCRLBs, for the number of RRHs $M$ ranging from $10$ to $100$. We notice that the RMSEs and BCRLBs decrease initially, followed by saturation. Firstly, this plot illustrates that it is beneficial from the location prediction point of view to choose massive MIMO systems over the conventional multiuser MIMO systems. Secondly, the BCRLB curve serves as a guideline to choose the number of RRHs for location prediction - for example, if we are operating with more than $50$ RRHs in the given area (due to coverage and/or throughput requirements), we can simply choose a subset of $50$ RRHs from the total number of RRHs for predicting the user locations and still be able to achieve the minimum possible RMSE performance.

\section{Conclusion}\label{sec_conclusion}
We have considered a Gaussian process regression (GP) framework to estimate user locations from their uplink received signal strength (RSS) data in a distributed massive multiple-input multiple-output (MIMO) system. Considering noise-free RSS data for training purposes and noisy RSS data for the test purposes, we have applied two GP methods for estimating the user locations, namely, the conventional GP (CGP) method and the moment-matching based numerical approximation GP (NaGP) method. We have firstly identified that the CGP method provides unrealistically small $2\sigma$ error-bars on the estimated locations because it naively treats the noisy test RSS data as noise-free. We have overcome this limitation in the proposed NaGP method because it learns from the statistical properties of the noise present in the test RSS data. We have also derived a Bayesian Cramer-Rao lower bound (BCRLB) on the achievable root-mean-squared error (RMSE) performance of the two GP methods under study and realized that the BCRLB can be obtained via simple linear algebraic operations on the predictive variances. Numerical studies  have provided few additional insights on the problem under study. Firstly, from the RSS-based location prediction point of view, it is beneficial to migrate from the conventional multiuser MIMO regime to the massive MIMO regime because we can achieve lower RMSE values. Secondly, it is possible to derive realistic $2\sigma$ error-bars on the estimated locations if we account for the noisy nature of the test RSS data, as is done by the NaGP method. Lastly, the BCRLB derived in this work can serve as a guideline to obtain the required number of BS antennas for user positioning.

Several exciting research directions may be pursued from the presented work. Firstly, we note that the reported RMSE values in this work are relatively high for use is wireless applications. This is essentially the cost we pay as a tradeoff for training the GP model with noise-free RSS data, which is easy to generate. We are currently working on analytical approximation GP methods which, apart from deriving realistic $2\sigma$ error-bars on the estimated locations, can also achieve lower RMSE than the CGP and NaGP methods \cite{gagp}. Secondly, for simplicity of exposition, we have modelled the $x$ and $y$ coordinates of the users as independent random variables. Better prediction performance may be achieved if the training procedure takes the correlation between the $x$ and $y$ coordinates of the user locations into account. 

\begin{appendix}
\subsection{Mathematical Formulae}
\begin{itemize}
	\item[(1)] [Conditioning a joint Gaussian distribution \cite{gpr_book} (pg. 200)] If $\mathbf{a}$ is a $W \times 1$ Gaussian random vector with $\mathbf{a} \sim \mathcal{N} (\mathbf{u}, \mathbf{A})$ and the random variables in $\mathbf{a}$ are partitioned into two sets $\mathbf{a}_{\zeta} = [[\mathbf{a}]_1 \, [\mathbf{a}]_2, \dots [\mathbf{a}]_{w}]^T \in \mathbb{R}^{w}$ and $\mathbf{a}_{\zeta'} = [[\mathbf{a}]_{w+1} \, [\mathbf{a}]_{w+2}, \dots [\mathbf{a}]_{W}]^T \in \mathbb{R}^{W-w}$  such that
	
	\begin{align} \label{eq_conditioning_formula}
	\begin{bmatrix}
	\mathbf{a}_{\zeta}\\
	\mathbf{a}_{\zeta'}
	\end{bmatrix}
	\sim  \mathcal{N}
	\begin{bmatrix}
	\begin{pmatrix}
	\mathbf{u}_{\zeta}\\
	\mathbf{u}_{\zeta'}
	\end{pmatrix}\!\!,
	\begin{pmatrix}
	\mathbf{A}_{\zeta \zeta} & \mathbf{A}_{\zeta \zeta'}\\
	\mathbf{A}_{\zeta \zeta'}^T & \mathbf{A}_{\zeta' \zeta'} 
	\end{pmatrix}
	\end{bmatrix}, 
	\end{align}
	then $\mathbf{a}_{\zeta} | \mathbf{a}_{\zeta'}$ and $\mathbf{a}_{\zeta'} | \mathbf{a}_{\zeta}$ are also Gaussian such that
	\begin{align}
	\mathbf{a}_{\zeta} | \mathbf{a}_{\zeta'} &\sim \mathcal{N} (\mathbf{u}_{\zeta} + \mathbf{A}_{\zeta \zeta'} \mathbf{A}_{\zeta' \zeta'}^{-1} (\mathbf{a}_{\zeta'} - \mathbf{u}_{\zeta'}), \mathbf{A}_{\zeta \zeta} - \nonumber\\
	&  \quad \quad \mathbf{A}_{\zeta \zeta'} \mathbf{A}_{\zeta' \zeta'}^{-1} \mathbf{A}_{\zeta \zeta'}^T), \nonumber \\
	\mathbf{a}_{\zeta'} | \mathbf{a}_{\zeta} &\sim \mathcal{N} (\mathbf{u}_{\zeta'} + \mathbf{A}_{\zeta \zeta'}^T \mathbf{A}_{\zeta \zeta}^{-1} (\mathbf{a}_{\zeta} - \mathbf{u}_{\zeta}), \mathbf{A}_{\zeta' \zeta'} - \nonumber \\
	& \quad \quad \mathbf{A}_{\zeta \zeta'}^T \mathbf{A}_{\zeta \zeta}^{-1} \mathbf{A}_{\zeta \zeta'}).
	\end{align}
	\item[(2)] [Product of Gaussian expressions] Let us consider three deterministic $W-$dimensional vectors $\mathbf{a}$, $\mathbf{u}$ and $\mathbf{u}_{0}$, and two $W \times W$ positive definite matrices $\mathbf{A}$ and $\mathbf{A}_0$. The product of Gaussian expressions $N (\mathbf{a}; \mathbf{u}, \mathbf{A}) $ and $N (\mathbf{a}; \mathbf{u}_0, \mathbf{A}_0)$ is then given by 
	\begin{equation} \label{gauss_prod}
	\begin{aligned} 
	& N (\mathbf{a}; \mathbf{u}, \mathbf{A}) N (\mathbf{a}; \mathbf{u}_0, \mathbf{A}_0)  \\
	& = N(\mathbf{u} ;\mathbf{u}_0 , \mathbf{A} + \mathbf{A}_0) N(\mathbf{a} ;\mathbf{u}_1 , \mathbf{A}_1),  \\ 
	& \text{ where } \mathbf{A}_1 = (\mathbf{A}^{-1} + \mathbf{A}_0^{-1})^{-1} \, \text{ and } \\ 
	&\quad \quad \quad \,\,\, \mathbf{u}_1 = \mathbf{A}_1 (\mathbf{A}^{-1} \mathbf{u} + \mathbf{A}_0^{-1} \mathbf{u}_0).
	\end{aligned}
	\end{equation}
	\item [(3)] [Covariance of a random vector] The covariance matrix $\mathbf{A}$ of a $W-$dimensional vector $\mathbf{a}$ has elements given by
	\begin{equation} \label{eq_cov_def}
	[\mathbf{A}]_{ww} = \mathbb{E}_{[\mathbf{a}]_w} (([\mathbf{a}]_w)^2) - (\mathbb{E}_{[\mathbf{a}]_w} ([\mathbf{a}]_w))^2, \, \forall w = 1, \dots, W.
	\end{equation}
	
\end{itemize}

\end{appendix}

\end{document}